# Dark Energy Stars and AdS/CFT

G. Chapline

Lawrence Livermore National Laboratory, Livermore, CA 94551

**Abstract**

The theory of dark energy stars illustrates how the behavior of matter near to certain kinds of quantum critical phase transitions can be given a geometrical interpretation by regarding the criticality tuning parameter as an "extra" dimension. In the case of a superfluid with vanishing speed of sound, the implied geometry resembles 5-dimensional anti-de-Sitter. In a dark energy star this geometry applies both *inside and outside* the horizon radius, so the AdS/CFT correspondence is consistent with the idea that the surface of a compact astrophysical object represents a quantum critical phase transition of space-time. The superfluid transition in a chiron gas, which was originally proposed as a theory of high temperature superconductivity, may provide an exact theory of this transition.

## 1. Introduction

Recently the use of a correspondence between conformal field theories and string theory in an anti-de-Sitter background to obtain estimates for quantum critical transport coefficients that are difficult to estimate directly has attracted considerable attention [1]. Part of the excitement surrounding these results derives from the hope that this "AdS/CFT correspondence" may also be interpreted as a theory of how classical space-time might emerge from an underlying quantum field theory. However, the AdS/CFT correspondence is a purely mathematical result which begs the question as to whether there actually is a close connection between the physics of quantum many body systems and the quantum physics of space-time. On the other hand an intimate connection between the behavior of quantum many body systems near to a quantum critical point and the physical nature of space-time does appear in the theory of dark energy stars. In this talk I wish to emphasize that this connection mirrors the AdS/CFT correspondence in a remarkable way. I will describe how the theory of quantum critical phase transitions can be given a geometrical interpretation by regarding the criticality tuning parameter as an "extra" dimension, and how this observation has resolved the long outstanding problem of the nature of space-time inside compact astrophysical objects formed by gravitational collapse.

The picture of gravitational collapse provided by classical general relativity (GR) cannot be completely correct because it conflicts with ordinary quantum mechanics during the final stages of collapse. In particular, the appearance of trapped surfaces makes it impossible to everywhere synchronize atomic clocks. There are other circumstances where GR doesn't allow for the global synchronization of clocks, e.g. rotating space-times, and so this is a generic defect of classical GR. As an alternative it has been suggested [2,3] that the vacuum state of space-time has off-diagonal order. This assumption implies [2] that during the final stages of the gravitational collapse the "squeezing" of the vacuum state increases dramatically. As a result it is expected that gravitational collapse of objects with masses greater than a few solar masses should lead to the formation of a compact object whose interior resembles de Sitter space-time with a large vacuum energy rather than the bizarre "black hole" space-times predicted by classical GR. A few years ago I introduced the name "*dark energy star*" for such objects [4].

In 2000 R. Laughlin and the author realized [5] that the mystery of gravitational collapse could be resolved if the surface of the compact object corresponds to a quantum critical phase transition of space-time. Quantum critical phase transitions have been observed in many kinds of condensed matter systems at low temperatures, and we suggested that the behavior of relativistic particles approaching a quantum critical surface of space-time could be surmised from the observed laboratory properties of real materials near to a quantum critical point. A simple thought experiment illustrates this possibility. Imagine a tall cylinder containing a superfluid with an equation of state such that a certain height the velocity of sound goes to zero (cf Fig.1). It is not hard to show [5] that the behavior of sound waves approaching the critical surface exactly mimics the behavior of relativistic particles approaching an event horizon.

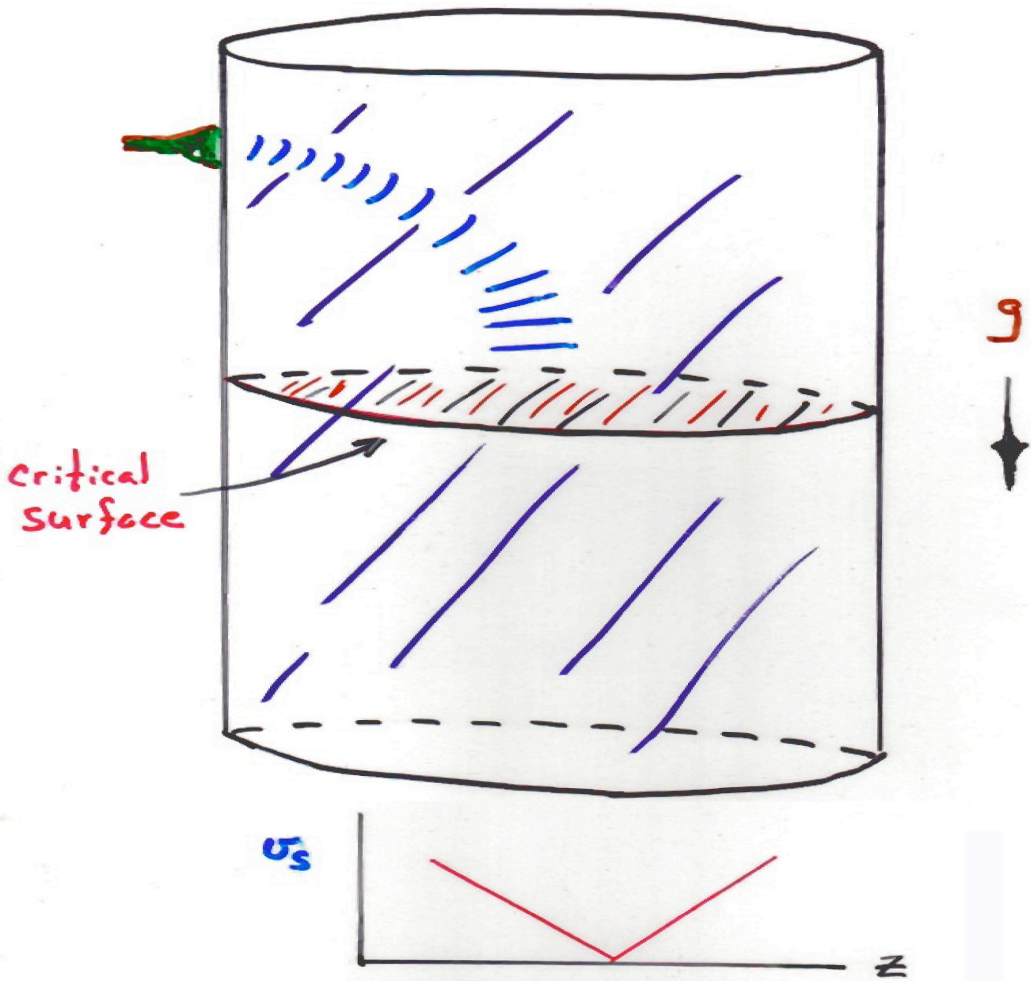

Fig. 1 Behavior of sound waves approaching a critical surface

On the basis the behavior of sound waves in our thought experiment we surmised that when matter approaches to within a distance $z^*$ from the surface where GR predicts there should be an event horizon, ordinary elementary particles morph into non-relativistic quantum particles. In particular, it follows from the Schrodinger equation for a superfluid that close to the critical surface the dispersion relation for small amplitude waves has the Bogoluibov form

$$\hbar\omega_q = \sqrt{(\hbar v_s q)^2 + \left(\frac{\hbar^2 q^2}{2m}\right)^2}, \quad (1)$$

where $h/mv_s$ is the coherence length. In the case of a compact object $v_s$ corresponds to $c(z/2R_g)$, where $R_g = 2GM/c^2$ is the Schwarzschild radius for a object with mass $M$. It follows from Eq. (1) that when relativistic particles approach to within a distance

$$z^* = R_g \sqrt{\hbar\omega/2mc^2} \quad (2)$$

from the Schwarzschild radius they will begin to behave like non-relativistic particles with mass m. This means that classical GR will break down at a distance from the Schwarzschild radius that depends on the energy of the particle. As developed previously [5] one of the consequences of the breakdown of a continuum superfluid description near to the surface is the appearance of dissipation and quantum critical fluctuations. The width of the critical layer in which quantum fluctuations are important depends on the frequency of the probe; this frequency dependence means that the scale factor for the space-time in which the quantum fluctuations live will depend on the macroscopic distance from the horizon radius.

The new picture that emerges for compact objects is that the interior space-time of the compact object looks like ordinary space-time except that that the vacuum energy is much larger than the cosmological vacuum energy. There is no space-time singularity in the interior. The time dilation factor for the interior metric is positive (in sharp contrast with the strange negative time dilation factor predicted by classical GR), but approaches zero as one approaches the event horizon surface (Fig. 2). Near the event horizon classical GR breaks down, and one needs a quantum theory of space-time to describe the transition from the exterior to the interior of the compact object. From the point of view of classical GR this transition layer must have unusual properties e.g the ability to support large stresses.

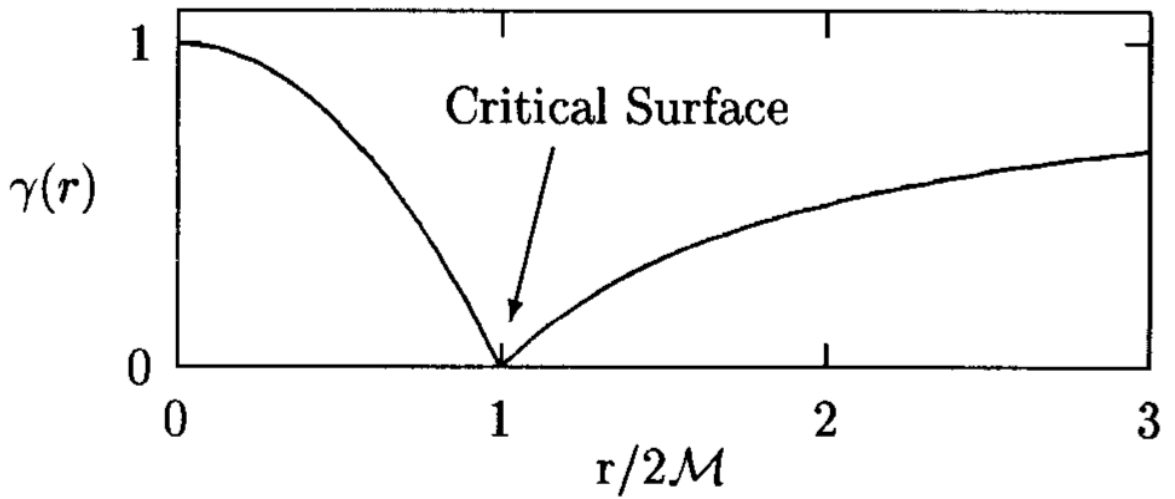

Fig. 2 Time dilation factor for a dark energy star

## 2. Anti-de-Sitter model for the surface of a dark energy star

According to the AdS/CFT correspondence a strongly coupled conformal field theory corresponds to classical anti-de-Sitter space where the spatial curvature is very small. A weakly coupled conformal field theory on the other hand corresponds to an anti-de-Sitter space with large curvature and "stringy" corrections to classical GR. Regarding the interior of a dark energy star as a superfluid with varying superfluid density reflects these correspondences in a remarkable way. The spatial curvature in the deep interior of the dark energy star is small and corresponds to a superfluid with strong particle corrrelations, while the surface of a dark energy star corresponds to a quantum critical point where superfluidity becomes weak and classical GR breaks down. This suggests that there may be a deep connection between the AdS/CFT correspondence and the physics of dark energy stars.

Actually the variation of the time dilation factor inside a dark energy star (cf Fig.2) strongly resembles the gravitational potential in an anti-de-Sitter space. One can make the connection with anti-de-Sitter space more dramatic by introducing an extra flat 3-dimensions as the setting for the local quantum degrees of freedom. If one demands that the metric satisfy the Einstein equations and that the scale factor for the extra 3-dimensions be consistent with eq. (2), then the metric in the composite space has the form

$$ds^2 = \frac{z^2}{R_H^2}\left(dt^2 - dx_i dx_i\right) - \frac{R_H^2}{z^2} dz^2 - R_H^2 d\Omega_2 \qquad , \qquad (3)$$

where $z$ is the distance from the surface of the dark energy star. The dark energy star radius $R_H$ plays the role of the radius of AdS space. In the case of a rotating dark energy star the radius satisfies [6] (in units where $8\pi G/c^2 = 1$):

$$R_H^2 + a^2 = \frac{\Lambda}{3} R_H^2 \qquad , \qquad (4)$$

where $\Lambda/3$ is the vacuum energy density inside the dark energy star and $a$ is the angular momentum per unit mass. If we replace the de Sitter radial coordinate $r$ with $R^2/r$, which is a more natural coordinate for anti-de-Sitter space, then this equation has the same form as the equation for the horizon radius of a black hole in anti-de-Sitter space.

In the dark energy star picture anti-de-Sitter geometry describes both the *interior and exterior* of a compact object near to its surface. Thus the AdS/CFT correspondence is consistent with the idea that the surface of a compact object represents a critical phase transition of the vacuum of space-time. As it happens a quantum theory of space-time was proposed some time ago [7] that apparently has just the right kind of phase transition.

## 3. Superfluid transition in a chiron gas

The essence of the dark energy star idea is that the vacuum of space-time corresponds to a superfluid, and that at the surface of a compact object the superfluid density increases by many orders of magnitude. Naturally one is led to wonder whether there are any examples of terrestrial superfluids that posses a quantum critical phase transition where the superfluid density increases dramatically as a result of changing some tuning parameter. It turns out that high temperature superconductivity may provide a good example of such a transition.

In high temperature superconductors the density of superconducting electrons increases dramatically at a certain density of, depending on the material, either electron or hole doping. The doping density where this happens depends on temperature in a characteristic way. As it happens a mathematical theory of high temperature superconductivity was proposed some years ago [8] that can explain the characteristic way in which the superconducting transition temperature in high temperature superconductors depends on doping density [9]. This theory generalizes a tight binding theory of the "parity anomaly" in graphene that had been proposed a few years earlier by Haldane [10]. A key feature of these models is the appearance of zero mass conduction states which replace the usual Fermi surface.

Our theory of high temperature superconductivity is based on the idea that in certain kinds of layered conductors with a low carrier density the conduction states can be described using a matrix non-linear Schrodinger equation with a non-abelian Chern-Simons gauge fields:

$$i\hbar \frac{\partial \Phi}{\partial t} = -\frac{1}{2m}D^2\Phi + e[A_0,\Phi] - g[[\Phi^*,\Phi],\Phi] , \qquad (5)$$

where the wave function $\Phi$ and potentials $A_0$, $A_i$ are N x N SU(N) matrices, and $D \equiv \nabla - i(e/\hbar c)[A,$ . N represents the number of layers, and the simplest solution to eq. (5) corresponds to assuming that the potentials lie in the adjoint representation of SU(N). The effective magnetic field $B_{eff} = \partial_x A_y - \partial_y A_x + [A_x, A_y]$ is a diagonal matrix:

$$B_{eff} = -\frac{e}{\kappa}[\Phi^*,\Phi] , \qquad (6)$$

so the effective magnetic field seen by charge carriers can vary from layer to layer. The in plane electric field $E_\alpha$ will also be a diagonal matrix:

$$E_\alpha = -\frac{1}{\kappa}\varepsilon_{\alpha\beta}j_\beta , \qquad (7)$$

where $j_\alpha = (\hbar/2mi)([\Phi^*,D_\alpha\Phi] - [D_\alpha\Phi^*,\Phi])$ is the in-plane current. Time independent analytic solutions to eq. (5-7) can be found for any value of N if the coupling constants e and κ satisfy the self-duality condition;

$$g = e^2\hbar/mc|\kappa|$$

These analytic solutions have zero energy and are either self-dual or anti-self-dual $D_\alpha\Phi = \pm i\varepsilon_{\alpha\beta}D_\beta\Phi$, depending on the sign of κ. Physically these two solutions correspond to having all the carrier spins be either up or down, and they can form a Kramers pair – which is the key to our understanding of high $T_c$ superconductivity.

In the limit N → ∞ the analytic solutions take a particularly simple form such that the effective magnetic field seen by the *j*th carrier is given by

$$B_j = \pm\frac{\hbar c}{e}\sum_k \nabla_k |X_j - X_k| , \qquad (8)$$

where $X \equiv (z, u)$ is now a 3-dimensional coordinate encoding both the position z = x+iy of a chiron within a layer and the height *u* of the layer. In this solution the vortex-like carriers present in the solution for a single layer have become monopole-like objects, which were christened "chirons" in ref. 8. These objects resemble polarons in that the electric charge is quasi-localized, but differ from polarons in that the charge localization also involves localized spin polarized currents. The wave function corresponding to (8) has the form

$$\Psi = f(w) \prod_{k>j}^{\infty} \left[ \frac{R_{jk} + U_{jk}}{R_{jk} - U_{jk}} \right]^{1/2}, \qquad (9)$$

where $R_{jk}^2 = U_{jk}^2 + 4(z_j - z_k)(\bar{z}_j - \bar{z}_k)$, $U_{jk} = u_j - u_k$, and $f$ is an entire function of the $\{\bar{z}_i\}$ in the self-dual case and $\{z_i\}$ in the anti-self-dual case. The wave function (9) resembles in some respects Laughlin's wave function for the fractional quantum Hall effect. However, our wave function describes a 3-dimensional system and, in contrast with the fractional quantum Hall effect, there are two distinct degenerate ground states corresponding to the self-dual and anti-self-dual solutions for eq. (5).

Writing the product on the rhs of eq. (9) as exp(S) defines an effective action for a chiron gas:

$$S = \frac{1}{2} \sum_j \ln \frac{R_j + u - u_j}{R_j - u + u_j}, \qquad (10)$$

where $R_j^2 = (u - u_j)^2 + 4(z - z_j)(\bar{z} - \bar{z}_j)$. The form of this effective action suggests a connection with the condensation of vortex and anti-vortex pairs in the 2-dimensional XY model. The configurations of XY spins implicated in the XY phase transition have the form:

$$\Theta(z) = \sum_i m_i \, \text{Im} \ln(z - z_i),$$

where the integer $m_i$ is the quantized circulation of the vortex (or anti-vortex if $m_i$ is negative) located at $z_i$. It is an elementary identity that the right hand side of (9) can be rewritten in the form

$$S = \sum_i \pm \tanh^{-1}\left( \frac{u - u_i}{R_i} \right), \qquad (11)$$

which is intriguingly similar in form to a configuration of 2-dimensional XY vortices.

In order to compare the behavior of gas of self-dual and anti-self-dual chirons with the behavior of the XY model we note that phase variations in a 2-dimensional condensate can be described by a partition function of the form

$$Z = \int_0^{2\pi} D\Theta \exp[-\frac{K}{2} \int d^2\xi \frac{\partial \Theta}{\partial \xi_i} \frac{\partial \Theta}{\partial \xi_i}], \qquad (12)$$

where $\Theta$ is a periodic coordinate whose period is $2\pi$ and K is a constant. It can be shown [11] that a discrete version of the 2-dimensional action (12) interpolates between the low and high temperature phases of the XY model. Indeed evaluating the exponential in (12) for a configuration of vortices yields the partition function for a 2-D Coulomb gas. On the other hand substituting the chiron effective action (11) into the exponential in (12) yields:

$$Z_c = \exp- \pi K \left[ \sum_{i \neq j} m_i m_j \ln \frac{R_{ij}}{|z_i - z_j|} \right], \quad (13)$$

where the "vorticity" $m = \pm 1$ means spin up or spin down, and the sum is restricted to equal numbers of up and down spins. When the average nearest neighbor distance between chirons within a layer is less than the spacing $U_{ij}$ between a particular pair of layers, the contribution of those layers to the partition function (13) resembles the partition function of a discrete 2-D Coulomb gas, with the inter-layer spacing playing the role of the lattice spacing in the discrete Coulomb gas. An important difference though is that only chirons in different layers attract one another. In the case of a 2-D Coulomb gas the KT transition would occur at the value $K = 2/\pi$, while the partition function for a trial ground state wave function which is simply a product of wave functions of the form (9) for spin up and spin down chirons corresponds to $K = 1$; i.e. just below the KT transition.

Evidently in the chiron theory the interlayer spacing $c$ serves as a regulator for a KT-like transition, in that the transition in our theory from a gas of free chirons to a condensate of chiron pairs should resemble a classical KT-like phase transition when the mean separation $d$ between chirons within each plane is less than say $2c$. If we identify $d = a/\sqrt{\delta}$, where $a$ is the lattice spacing in the plane and $\delta$ is the doping, then $d = 2c$ would correspond in the case of the high temperature superconductor LSCO to $\delta = .08$. In the condensate phase of the chiron gas each spin up or spin down chiron is paired with an opposite spin chiron in a neighboring layer, so the superfluid density is $(\frac{1}{2})d^{-2}$. However, the formation of a condensate in our theory is not exactly a KT phase transition because the potentials between chirons are not simple logarithms. Instead, the condensation transition in our model will be a smooth cross-over, and only closely resemble a KT phase transition when $d < 2c$. However the characteristic temperature where this cross-over takes place can be estimated in a fashion analogous to the original reasoning of Kosterlitz and Thouless [12] by comparing the effective potential for spin up and spin down chirons with the 2-dimensional positional entropy of the chirons. Keeping just the contribution of the nearest and next nearest layers in (13) and comparing with the entropy of the chirons located on a lattice with spacing $a$ leads to the following estimate for the cross-over temperature, which should be applicable for low densities of chirons such that $d > c$:

$$T_c = \frac{\pi \hbar^2}{4 m^* d^2} \frac{0.5 \ln(1 + \frac{c^2}{d^2}) + 0.5 \ln(1 + \frac{(2c)^2}{d^2})}{\ln(d^2/a^2)} . \quad (14)$$

A comparison of the characteristic cross-over temperature predicted by this relation using the values $a = 3.8$, $m^* = 4 m_e$, and $c/a = 1.7$ appropriate to LCSO with observed transition temperatures in underdoped LSCO is shown in Fig 3 (see ref. 9 for details). The prediction (14) for the transition temperature will fail at small values of $d$ because ordinary electrostatic screening will hide the Chern-Simons attraction between chirons.

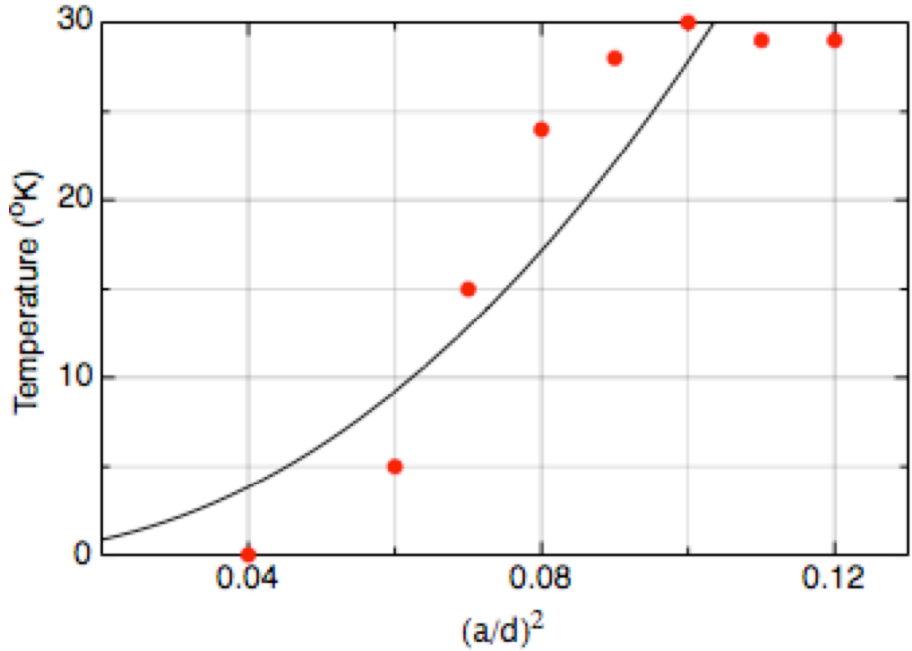

Fig 3. Predicted superfluid transition temperature in LSCO

In summary, the chiron theory of high Tc superconductivity yields a quantum phase transition where at low temperatures the superfluid density rapidly increase at a critical carrier density. As pointed out in ref. 7 the chiron theory can also be interpreted as an "ambi-twistor" theory of space-time. Thus it is quite possible that a superfluid transition similar to that observed in high temperature superconductors occurs at the surface of compact astrophysical objects.

## 4. Epilogue

One amusing aspect of the simple superfluid model for a dark energy star introduced in ref. 5 is the occurrence of string-like surface excitations. These excitations can be found be solving the non-linear Schrodinger equation with a varying speed of sound. If we assume that the velocity of sound depends linearly on the height above the surface one obtains the resonances in sound wave reflectivity shown in Fig 4. The resonant frequencies depend on the momentum $Q$ parallel to the surface and the slope $1/\tau_0$ of the speed of sound vs. height; and for large Q become harmonic with fundamental frequency $\sqrt{2h}/\tau_0$ .

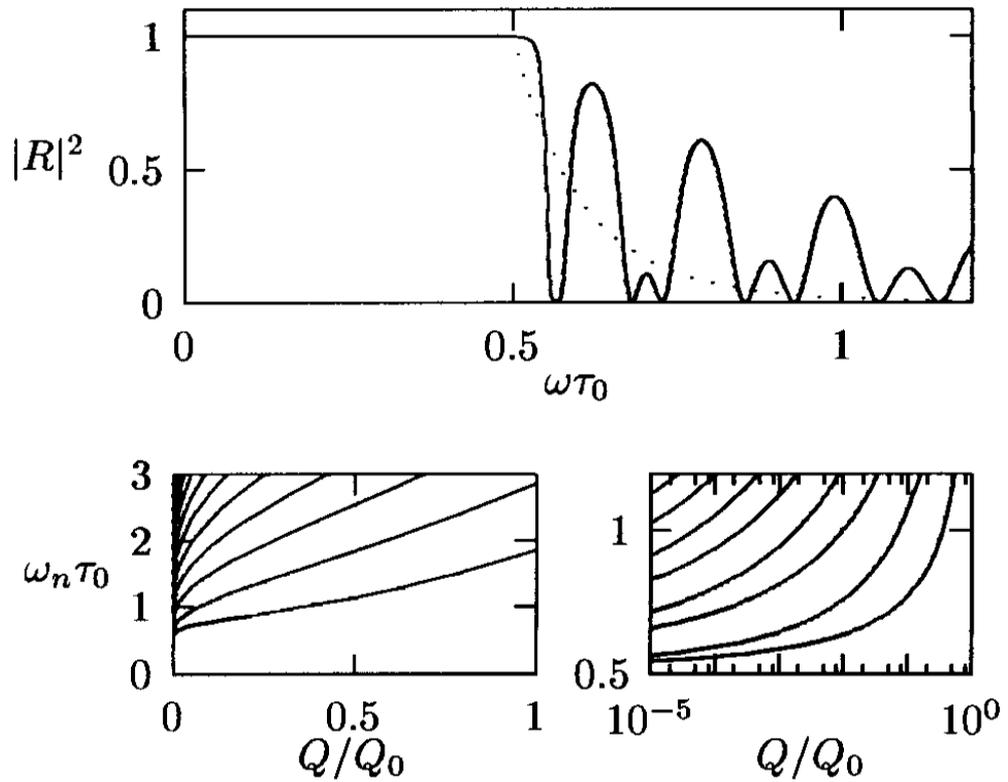

Fig. 4 Surface resonances for a dark energy star

One is tempted to imagine that these string-like collective excitations form an alternative description of the critical surface layer. In any case the theory of dark energy stars brings us full circle back to the origin of string theory as a theory of wave propagation around a spherical object. String theory originated with the interest of Sommerfeld and Watson in the problem of over the horizon radio reception. Their work was later extended to a theory of radio propagation around the earth by van der Pol and Bremmer [13].

Acknowledgments

The author is very grateful for conversations with Pawel Mazur and Jan Zaanen, This work was performed under the auspices of the U.S. Department of Energy by Lawrence Livermore National Laboratory under Contract DE-AC52-07NA27344.

References

1. J. Zaanen, Nature **448**, 1000 (2007).
2. G. Chapline, in *Foundations of Quantum Mechanics*, ed. by T. Black et. al. (Singapore, World Scientific), 1992).
3. P. Mazur, Acta Phys.Polon. B27, 1849 (1996), (hep-th/9603014).
4. G. Chapline, Proceedings of the 22[nd] Texas Conference on Relativistic Astrophysics (Stanford University 2005).
5. Chapline, G., Hohlfeld E., Laughlin, R. & Santiago, D. 2001, Phil. Mag. B, 81, 235 .
6. G. Chapline and P. Marecki, astro-ph: 0809.1115v2.

7. G. Chapline, Mod. Phys. Lett A7, 1959 (1992).
8. G. Chapline and K. Yamagishi, Phys. Rev. Lett., **66**, 304 (1991).
9. G. Chapline, Phil Mag **88,** 1227 (2008).
10. F. D. Haldane, Phys. Rev. Lett. **61,** 2015 (1988).
11. G. Chapline and F. Klinkhamer, Mod. Phys Lett. A 11, 1063 (1989).
12. J. M. Kosterlitz and D. J. Thouless, J. Phys. C 6, 1181 (1973).
13. B. Van der Pol and H. Bremmer, Phil Mag **24,** 141 (1937).